\newif\ifmnras
\mnrastrue
\ifmnras
   	\documentclass[apj,numberedappendix]{emulateapj}
\else
	\documentclass[apj,numberedappendix]{emulateapj}
\fi

\usepackage{graphics,epsf}
\usepackage{amsmath}                
\usepackage{amsfonts}               
\usepackage{amssymb}                
\usepackage{epsfig}                 
\usepackage{graphicx}

\def \AU{~\rm{AU}}

\def \Gyr{~\rm{Gyr}}

\ifmnras
	\def \aap{A\&A}
	
	\def \apj{ApJ}
	\def \apjl{ApJ}
	\def \apjs{ApJS}

	\def \mnras{MNRAS}

\fi

\usepackage{xcolor}
\definecolor{redak}{rgb}{0.9,0.15,0.05}

\begin{document}

\ifmnras

\title{The Class of Jsolated Stars and Luminous Planetary Nebulae in old stellar populations}
\author{ Efrat Sabach\altaffilmark{1} \& Noam Soker\altaffilmark{1,2}}

\altaffiltext{1}{Department of Physics, Technion -- Israel Institute of Technology, Haifa 32000, Israel; aldanag@campus.technion.ac.il; efrats@physics.technion.ac.il; soker@physics.technion.ac.il}
\altaffiltext{2}{Guangdong Technion Israel Institute of Technology, Shantou, Guangdong Province, China}

\begin{abstract}
We suggest that stars whose angular momentum (J) does not increase by a
companion, star or planet, along their post-main sequence evolution
have much lower mass loss rates along their giant branches.
Their classification to a separate group can bring insight on their late
evolution stages. 
We here term these {\it Jsolated stars}.
We argue that the mass loss rate of Jsolated stars is poorly determined
because the mass loss rate expressions on the giant branches are empirically
based on samples containing stars that experience strong binary interaction, with
stellar or sub-stellar companions, e.g., planetary nebula (PN) progenitors.
We use our earlier claim for a low mass loss rate of asymptotic giant branch (AGB) stars that are not spun-up by a stellar or substellar companion to show that we can account for the enigmatic finding that the brightest PNe in old stellar populations reach the same luminosity as the brightest PNe in young populations. It is quite likely that the best solution to the existence of bright PNe in old stellar populations is the combination of higher AGB luminosities, as obtained in some new stellar models, and the lower mass loss rates invoked here.
\end{abstract}


\section{INTRODUCTION}
\label{sec:intro}

Detailed observations over the years have strengthened the notion that the
mass loss process of evolved stars is highly variable.
Examples include the huge mass loss rate of luminous blue variable (LBV)
stars that experience major eruptions, such as $\eta$ Carinae
(e.g., \citealt{SmithOwocki2006}), pre-outbursts of some core
collapse supernovae (CCSNe; e.g., \citealt{Mauerhanetal2014, Grahametal2014,
SvirskiNakar2014, Tartagliaetal2016, Ofeketal2016}),
and the dense shells and lobes of many planetary nebulae (PNe),
as evident from hundreds of images (e.g., \citealt{Balick1987, Chuetal1987,
CorradiSchwarz1995, Manchadoetal1996, SahaiTrauger1998, Parkeretal2016}).
To those we can add the three-rings around the progenitor of SN~1987A
\citep{Burrowsetal1995}, and similar blue stars such as SBW1
\citep{Smithetal2013}, as well as progenitors of stripped CCSNe, such as SN Ib
(e.g., \citealt{Kangasetal2017}).

An interacting stellar companion can deposit energy and angular momentum to
the primary mass-losing star. Interaction with stellar companions outside the envelope of the primary star, mainly by tidal spin-up, suggests that in many cases angular momentum plays a larger role in increasing the mass loss rate than energy deposition does \citep{Soker2004}. 
Evolved red giant branch (RGB) and asymptotic giant branch (AGB)
stars can acquire a large amount of angular momentum by swallowing planets
(e.g., \citealt{Soker1996, Carlbergetal2009, VillaverLivio2009, MustillVillaver2012,
NordhausSpiegel2013, GarciaSeguraetal2014, Staffetal2016, AguileraGomezetal2016}).
In those cases deposition of angular momentum is more significant than
deposition of energy, e.g., for the operation of a dynamo in the envelope
of the giant star (e.g., \citealt{NordhausBlackman2006}).
Stars whose angular momentum $J$ does not increase in their post-main
sequence evolution experience a different mass loss history than those that
suffer interaction with stars, brown dwarfs, and planets.

A star that along its entire evolution does not acquire angular momentum
from a stellar companion or a sub-stellar companion, or that the angular
momentum it acquires $J_{\rm dep}$ is less than a fraction $\beta_j$ of the
maximum value it can have on the main sequence $J_{\rm MS, max}$,
is here termed a J isolated star, or {\it Jsolated star},
\begin{equation}
J_{\rm dep} \le \beta_j J_{\rm MS, max} \qquad {\rm for~ a~ Jsolated~ star}\footnote{ We define the maximum angular momentum on the main sequence to be the value such that the star does not reach its break-up angular velocity, $\omega_b=(G M/R^3)^{1/2}$, where $M$ and $R$ are the mass and radius of the star, under the assumption that it maintains its spherical structure, $J_{\rm MS, max}=\omega_b I_{\rm MS}$, where $I_{\rm MS}$ is the moment of inertia of the (spherical) star.} . 
\label{eq:jsolated}
\end{equation}
At this point there is no accurate theory for the mass loss rate of red giant
stars, and we cannot determine the exact value of $\beta_J$.
In this preliminary study of the properties of Jsolated stars, we take a crude
estimate of $\beta_J \simeq 0.1-1$. 
In any case, the exact value of $\beta_J$ has no real influence on the
conclusions reached in this study. 

A large fraction of stars with zero-age main sequence mass of
$M_{\rm ZAMS} > 1 M_\odot$ are in close binary systems, or harbour planetary
systems (e.g., \citealt{Bowleretal2010}).
These are sufficient to account for observed PNe \citep{DeMarcoSoker2011},
as it seems that most PNe result from binary interaction
\citep{DeMarcoMoe2005, SokerSubag2005, MoeDeMarco2006}.
Namely, Jsolated stars form no PNe, or at most they form spherical and very faint
PNe (also termed hidden PNe).   
When massive stars are considered, the fraction of stars that will experience post-main
sequence interaction increases (e.g., \citealt{MoeDiStefano2015}).
This implies that the fraction of Jsolated stars steeply decreases with increasing
initial stellar mass. 

The above considerations and the list of evolved stars that result from binary
interaction suggest that as much as our understanding of the mass loss process
from evolved interacting stars in binary (and triple) systems is poor, our
knowledge of the mass loss rate from Jsolated stars is poorer.
It should be noted that the definition of the Jsolated group is based on a
speculation that if a star is not spun up during its post-MS evolution, its mass
loss rate is much lower than the average mass loss rate observed.
We here propose that the mass loss rate observed in
evolved stars, such as
progenitors of PNe, is a result of a binary interaction.
Namely, the fitting formulae of the mass loss rates from RGB and AGB stars are actually
applicable to stars that experienced binary interaction, with stellar or sub-stellar
objects, during their post-main sequence evolution (non-Jsolated stars).
We further propose that the mass loss rate on the RGB and AGB of Jsolated stars
is much lower than what these fitting formulae give.
Jsolated star, and their reduced mass loss, might solve two puzzles; First is the puzzle of shaping mildly elliptical PNe, second is the puzzle of bright PNe in old stellar populations.

In \cite{SabachSoker2018} we have presented the idea of a reduced mass loss and its implications on solar-type stars.
We have shown that stars with a reduced mass loss rate can expand to larger radii than usually assumed, hence are much more likely to interact with planets on the upper AGB.
If a star does swallow a planet, on the upper AGB, then strictly speaking it is not a Jsolated star any more.
This might account for many PNe that are shaped by planets as suggested by \cite{Soker1996}.
We also found out that such stars reach much higher luminosities at the end of their AGB phase.
We here set the term {\it Jsolated stars} and further use this assumption to study not only shaping of PNe and ejecting a denser nebula, but also to address the luminosity of the central star.
We examine the late evolution in detail, aiming to explain the intriguing puzzle
of the planetary nebula luminosity function (PNLF)
(e.g., \citealt{Ciardullo2010,  Ciardulloetal1989, Ciardulloetal2005, Davisetal2018, Jacoby1989, vandeSteeneetal2006}).

The bright-end cutoff of the PNLF in [O~III]~$\lambda$~5007 has no dependence  
on the age or metallicity of the stellar population, meaning that stars in old stellar populations, where $M_{\rm ZAMS}\simeq 1-1.2M_\odot$, have their brightest [O~III]~$\lambda$~5007 PNe as in young stellar populations.
To reach such bright PNe the central star should reach luminosities of $\simeq 5000 L_\odot$ and higher in order to ionize such a bright nebula.
Such high luminosities are not reached with the traditional stellar
evolution calculations.
We here propose that the low mass loss rate of Jsolated stars with a late engulfment of a low mass companion (planet, or a brown dwarf, or a low mass star), account for the brightest PNe in old stellar populations
and might answer this puzzle.

There are some earlier studies that attempted to explain the constant bright end of the PNLF (e.g., \citealt{Richeretal1997}).  \cite{Ciardulloetal2005} proposed that the brightest PNe in old stellar populations evolve from blue-stragglers, namely, two lower mass stars that merged and formed a sufficiently massive star to form a bright PN. 
\cite{Soker2006a} proposed that the bright PNe are actually symbiotic nebulae, namely, a white dwarf companion to a giant star ionizes the nebula blown by the giant (for a debate on this scenario see \citealt{Ciardullo2006} and  \citealt{FrankowskiSoker2009}). 
\cite{FrankowskiSoker2009} list these two explanations, as well as one that considers a lower mass loss rate due to low metallicity in old stellar populations (see discussion in \citealt{Mendezetal2008}). In the present paper we attribute the lower mass loss rate to the absence of binary interaction until the upper AGB. 

There are also numerical studies that obtained high luminosities along the AGB but did not address the puzzle of the constant bright end of the PNLF (e.g.,  \citealt{Karakas2014, Bertolami2016}). 
In a recent paper \cite{Venturaetal2018} simulated the evolution of stars of different masses. Their model of a star with an initial mass of $1.25 M_\odot$ and a metallicity of $Z=0.014$ reaches a maximum AGB luminosity of $6530 L_\odot$. 
This is sufficient to ionize a bright PN. However, their model of a $1 M_\odot$ star reaches a maximum luminosity  of only $3800 L_\odot$, too low to account for the brightest PNe in very old stellar populations. 
Neither \cite{Karakas2014} nor \cite{Venturaetal2018} address the PNLF puzzle, and so we cannot tell whether their evolutionary routes can account for the constant bright end of the PNLF.
For example, in addition to high luminosities a fast evolution of the post-AGB is required. We attribute the fast post-AGB evolution of the brightest PNe in old stellar populations to an interaction with a low mass companion (a planet, a brown dwarf, or a low mass main sequence star). 
It will be interesting to use these evolutionary tracks to examine the PNLF. However, we are using the evolutionary stellar code MESA, with which we cannot reach the high luminosities they reach, unless we reduce the mass loss rate on the giant branches.  
As we will explain later, it is possible that both approaches, that of the new stellar evolutionary codes (e.g., \citealt{Venturaetal2018}) and a lower mass loss rate that is followed by the engulfment of a planet, as we suggest here, will have to be combined to obtain a satisfactory explanation for the constant bright end of the PNLF.

The new evolutionary calculations of Miller Bertolami (\citealt{Bertolamietal2008, Bertolami2016, Gesickietal2017, Reindletal2017}) have been used to address the bright end of the PNLF. 
\cite{Mendez2017} shows a fit to the PNLF of NGC 4697 based on the \cite{Bertolami2016} tracks, yet notes that it is not a full solution to the PNLF puzzle. In a very recent paper \cite{Gesickietal2018} study the bright end of the PNLF. They can account for most cases, but not for stellar populations older than about $7 \Gyr$. Again, we suggest that the usage of both these new stellar evolution simulations and our proposed reduced mass loss rate will solve the full puzzle of the bright end of the PNLF, but in the present study we use the stellar evolutionary code MESA to reveal the role of reduced mass loss.

In section \ref{sec:massloss} we elaborate on the mass loss rate. In section \ref{sec:method} we study the late evolution of Jsolated stars and present our results.
In section \ref{sec:discussion&summary} we discuss the
implications of our treatment of Jsolated stars, mainly regarding the PNLF.
 
\section{REDUCED MASS LOSS RATE}
\label{sec:massloss}
The empirical mass loss formula for red giant stars of \cite{Reimers1975}
is commonly expressed as
\begin{equation}
\dot{M}=\eta\times4\times10^{-13}LM^{-1}R,
\label{eq:Reimers}
\end{equation}
where $M$, $L$ and $R$ are in solar units and have their usual meaning,
and $\eta$ is the scaling parameter for the mass loss rate efficiency
set by observational constraints.
Here we do not go into the mass loss prescription or mechanism as this
is a topic of much review (e.g., \citealt{Lafon&Berruyer1991, SchroderCuntz2005}).
We only study the effects of mass loss reduction on stellar evolution
under the assumption that Jsolated stars experience a much
lower mass loss rate than non-Jsolated stars.

We change the mass loss rate efficiency parameter $\eta$ from $\eta=0.5$,
as commonly taken for solar-type stars, and reduce it as low as 0.05.
We point out that several works have studied the value of the mass loss rate
in RGB stars, such as \cite{McDonaldZijlstra2015} that find a  median value
of $\eta=0.477$ from horizontal branch (HB) morphology in globular clusters.
Our claim that if a star is not spun up during its post-MS evolution its mass
loss rate is much lower than the average value, together with
the values found for $\eta$, such as by
\cite{McDonaldZijlstra2015}, implies that most stars do suffer some kind of
interaction on their RGB.
Such interaction can be a tidal interaction with a companion at several AUs, or a common envelope
with a low mass companion.
As planets are not expected in globular clusters, the finding of \cite{McDonaldZijlstra2015}
might seem at first to rule out our assumption.
However the study of \cite{McDonaldZijlstra2015} cannot account
for all observations, such as bright PNe in old stellar populations.
If Jsolated stars are a minority, then their finding does
not contradict our assumption.
\cite{Miglioetal2012} find that the RGB mass loss rate parameter might be as low as 
$\eta=0.1$, i.e., lower than typically taken, for the old metal-rich cluster NGC 6791.
We differ in that we conduct a systematic comparison, and attribute the low
mass loss rate to Jsolated stars. 
In light of the problem to explain some properties of bright PNe in old populations
and the upper mass of white dwarfs, we claim that our speculative definition of Jsolated stars
has a merit, and so we conduct the study to follow.

We note that our idea that planet interacts with low mass AGB stars only on the upper AGB is compatible with the observations that many elliptical PNe have spherical and faint halos (e.g.  \citealt{Corradietal2003}). As single stars lose most of their angular momentum before reaching the upper AGB (e.g., \citealt{Soker2006b}), they will eject a faint and spherical halo. Only when interacting with a low mass companion on the upper AGB they terminate their AGB with a higher mass loss rate phase and an aspherical one (e.g., \citealt{Soker2000}). 

\section{NUMERICAL METHOD AND RESULTS}
\label{sec:method}
To study some effects of mass loss we conduct stellar evolution simulations using 
the Modules for Experiments in Stellar Astrophysics (MESA),
version 9575 (\citealt{Paxtonetal2011, Paxtonetal2013, Paxtonetal2015}).
We calculate stellar evolution from zero age main sequence until
the formation of a white dwarf for four stars in the mass range of $1-2M_\odot$ while varying the mass loss rate,
with solar metallicity 
of $Z=0.02$ (\citealt{vonSteigerZurbuchen2016} find $Z_\odot=0.0196$).
We point out that as there is an ongoing debate as to the exact
value of the solar metallicity, e.g., \cite{Vagnozzietal2017}, 
we also ran our simulations for a lower metallicity of $Z=0.014$
and derived similar results.

We study stellar models with four initial masses, and for each compare
the evolution with the commonly used mass loss rate efficiency parameter
to the evolution with reduced mass loss.
The typical value of the Reimers parameter for solar type stars is $\approx 0.5$ (e.g., \citealt{Guoetal2017}).
Here we treat the "typical" evolution of Sun-like stars and specifically the mass loss
mechanisms to be as taken in MESA,
where the \cite{Reimers1975} mass loss prescription is taken for the RGB and the prescription of
\cite{Bloecker1995} is taken for the AGB.
We do not focus on one certain value for Jsolated stars but rather examine a range of values of $\eta$ and the resulting evolution.
We postulate that Jsolated stars have a much lower mass loss efficiency parameter of $\eta\la0.1$.

In Figure \ref{fig:evolution} we present the evolution for the two lower masses that are relevant to old stellar populations, with 
initial masses of (a) $M_i=1M_\odot$, and (b) $M_i=1.2M_\odot$.
We examined the evolution with the commonly used mass loss rate efficiency parameter, $\eta=0.5$, and with several reduced values: 
$\eta=0.35$, $\eta=0.25$, $\eta=0.15$, $\eta=0.07$, and $\eta=0.05$.
We present the evolution of the stellar radius in the upper panel.
This is similar to our earlier analysis in \cite{SabachSoker2018} yet here we do not
focus on a single value for the reduced mass loss rate of Sun-like stars, but rather examine
a range of parameters and their result.
Moreover, we look at the general case
of a Jupiter-like stellar companion rather than at specific observed exoplanets. 

In the lower panels we examine the ratio of the stellar radius $R_\ast$ to the orbital separation $a$ of a Jupiter mass planet that has an initial (zero age main sequence) orbital separation of $a_i= 3\AU$.
Similar to our treatment in \cite{SabachSoker2018}, we here
calculate the evolution of the planet's orbital separation under the influence of the stellar mass loss alone, ignoring the effects of tidal interaction. Namely, we follow the mass loss from the star and change the orbital separation to conserve orbital angular momentum. Large mass loss episodes take place on the RGB and on the AGB, and in these phases the orbital separation increases. We follow the ratio $R_\ast/a$ to determine the possibility for planet engulfment.
As the ratio on the tip of the AGB becomes $R_{\rm max}/a \simeq 0.5$,
we expect that tidal interaction will bring a planet of mass
$m_p \ga M_{\rm Jupiter}$ into the envelope at the tip of the AGB
\citep{Soker1996} and will likely form an elliptical PN \citep{DeMarcoSoker2011}.
We find that already for $\eta=0.15$ planet engulfment will occur, hinting to a much more robust result than we have obtained in our first paper \citep{SabachSoker2018}.
\begin{figure*}
\centering
{(a)}
\includegraphics
[trim= 3cm 6cm 4cm 4cm,clip=true,width=0.45\textwidth]
{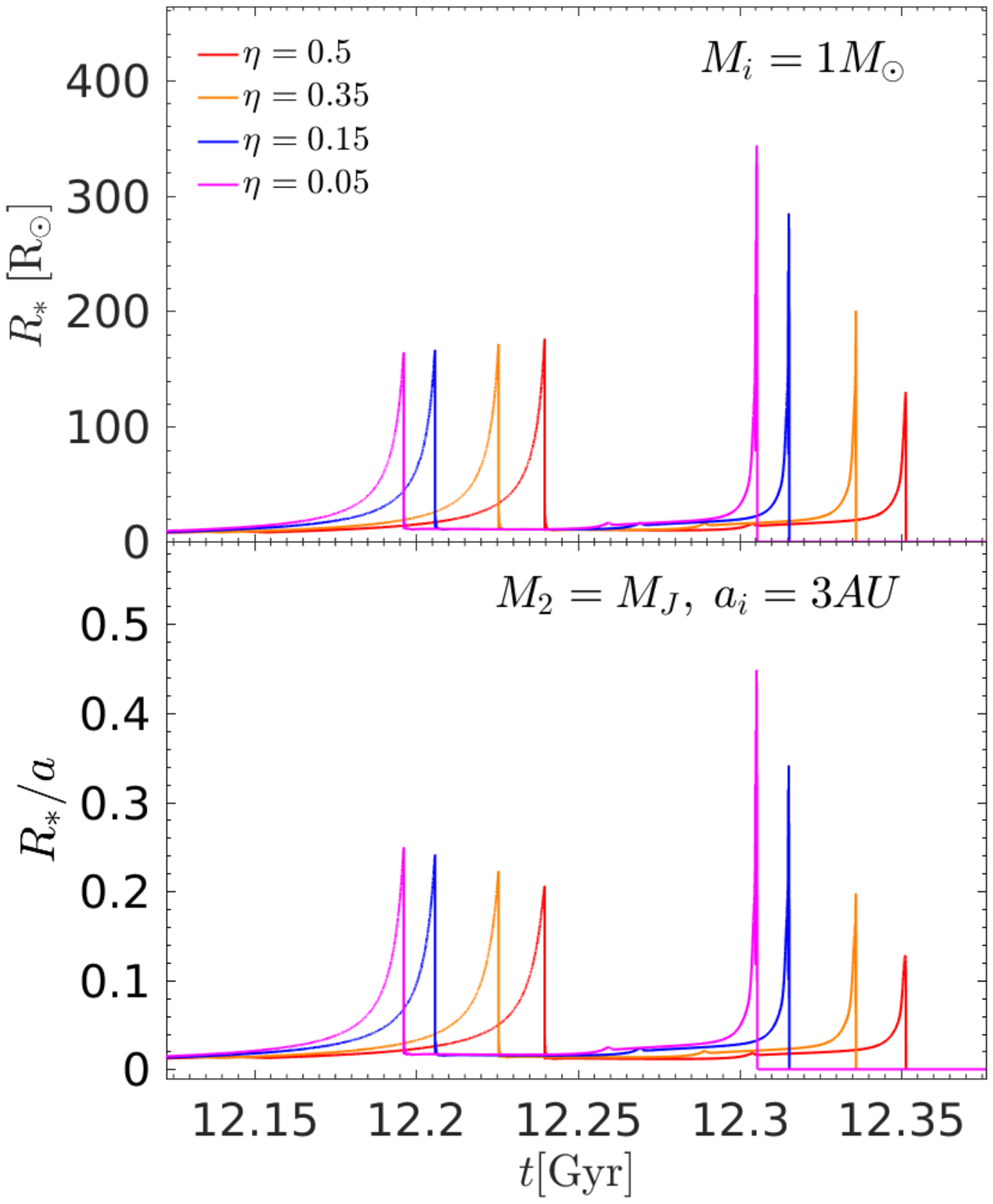}
{(b)}
\includegraphics
[trim= 3cm 6cm 4cm 4cm,clip=true,width=0.45\textwidth]
{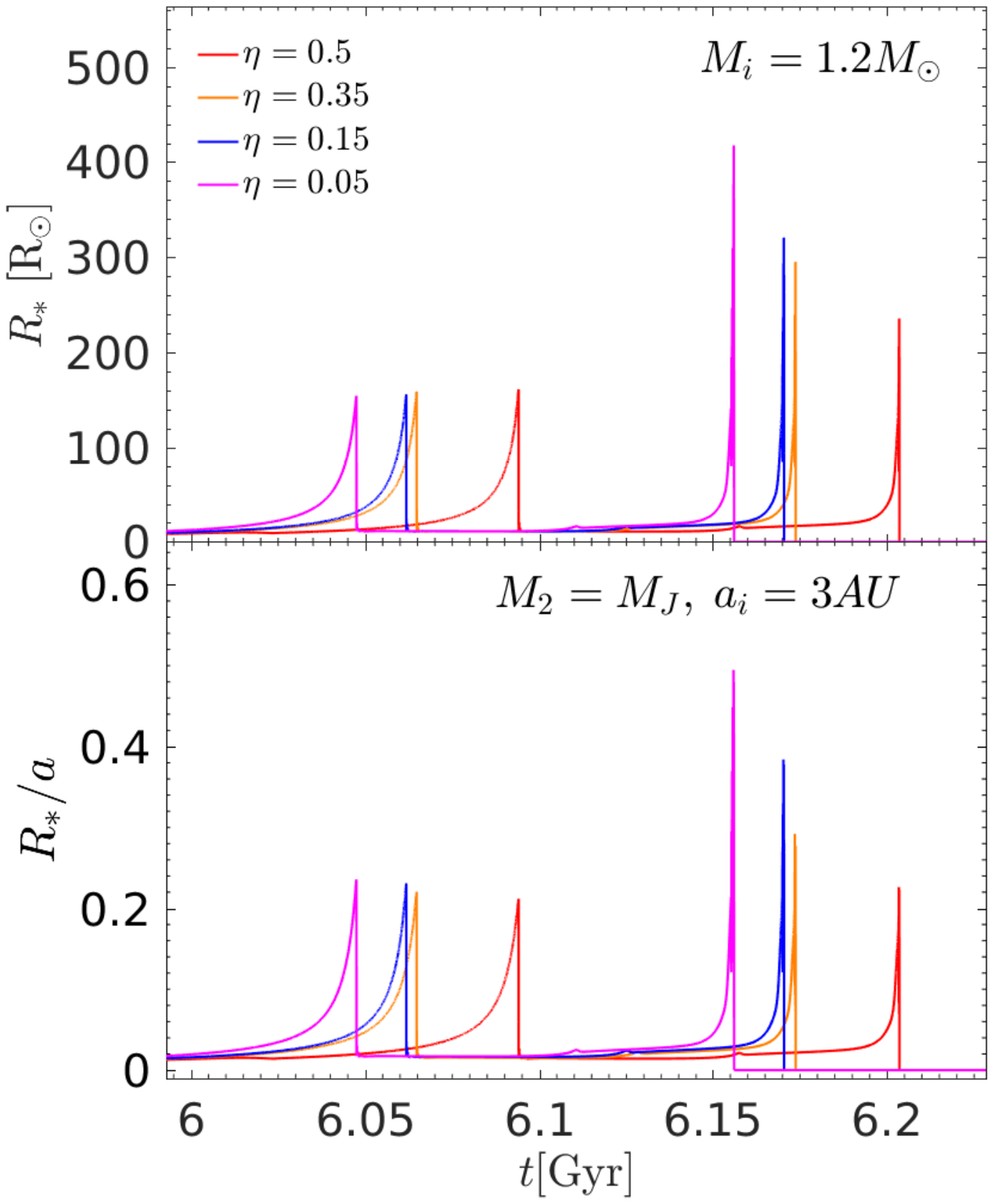}
\caption
{ The evolution
of two stellar models with initial masses of (a) $M_i=1M_\odot$, left panel, and (b) $1.2M_\odot$, right panel, calculated with MESA.
We calculate the evolution from zero age main sequence until the formation
of a white dwarf, yet the plots focus on the giant phases only.
The upper panels show the stellar radius $R_\ast$.
The lower panels show the ratio of the stellar radius to the separation of a Jupiter mass planet, $R_\ast/a$, with an initial semi-major axis of $a_i=3 \AU$. 
This calculation includes stellar mass loss but not tidal interaction.
We examined several mass loss rate efficiency parameters, and present 
here the results for 
$\eta=0.5$ (red), $\eta=0.35$ (orange), $\eta=0.15$ (blue) and $\eta=0.05$ (magenta).
The two giant phases appear as two peaks for each line (different color), the first one on the RGB and the second one on the AGB.
}
\label{fig:evolution}
\end{figure*}

In Figures \ref{fig:AGB_1M}-\ref{fig:AGB_2M} we present the mass, the luminosity and the radius on the upper AGB of our four stellar models. To account for the bright end of the PNLF the low mass stars must reach luminosities of $\ga 5000L_\odot$. We present the two higher masses for comparison.
It is apparent that for solar-like stars these values are not reached with the commonly used mass loss rate parameter and with the stellar evolutionary code MESA, yet as we reduce the mass loss parameter to about $\eta=0.15$ and below the needed values of the luminosity to account for the PNLF cut-off are reached.
 
We summarize the important results in Table \ref{tab:results}.
Notice the values of the maximum AGB radius (column 3) and the maximum AGB luminosity (column 4) are approximate values taken over the final AGB phase, since the radius and luminosity vary non-monotonically along the AGB. The correct values that should be used are the values the AGB star has at the time it engulfs its low mass companion.
Under our assumption the interaction with a planet (or a brown dwarf, or a low mass main sequence star) leads to a higher mass loss rate and a rapid post-AGB evolution, both of which are required in addition to a high stellar luminosity for the formation of a bright PN. Since we do not follow a specific system to determine the exact time of engulfment, we use the representative values on the final AGB phase.
\begin{figure}
\centering
\includegraphics
[trim= 3cm 1cm 4.6cm 1cm,clip=true,width=0.45\textwidth]
{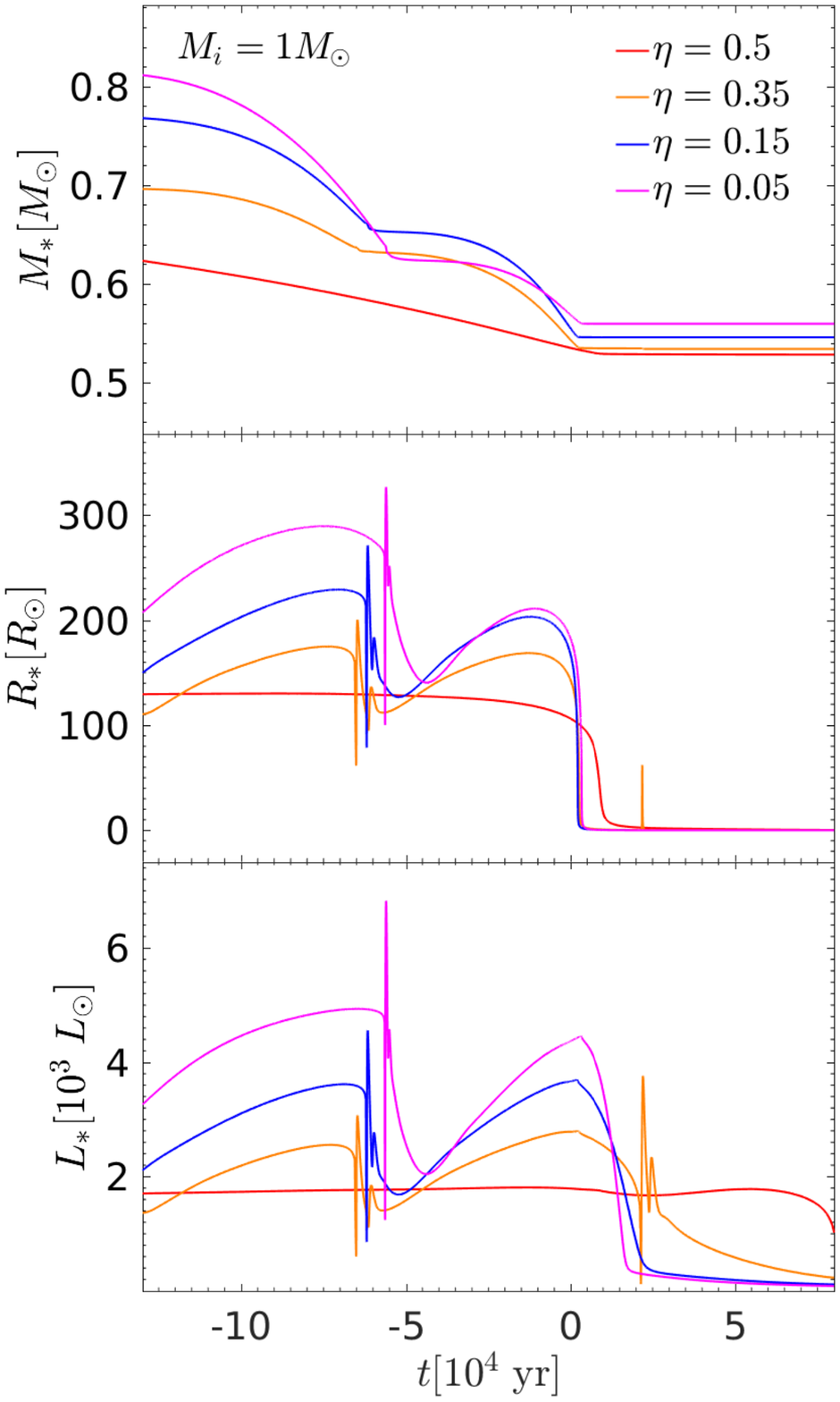}
\caption
{Evolution of an $M_i=1 M_\odot$ star during the
final  $\simeq2\times10^5\rm yr$ of the AGB phase calculated with MESA.
We examined several mass loss rate efficiency parameters, and present 
here the results for 
$\eta=0.5$ (red), $\eta=0.35$ (orange), $\eta=0.15$ (blue) and $\eta=0.05$ (magenta).}
\label{fig:AGB_1M}
\end{figure}
\begin{figure}
\centering
\includegraphics
[trim= 3cm 1cm 4.6cm 1cm,clip=true,width=0.45\textwidth]
{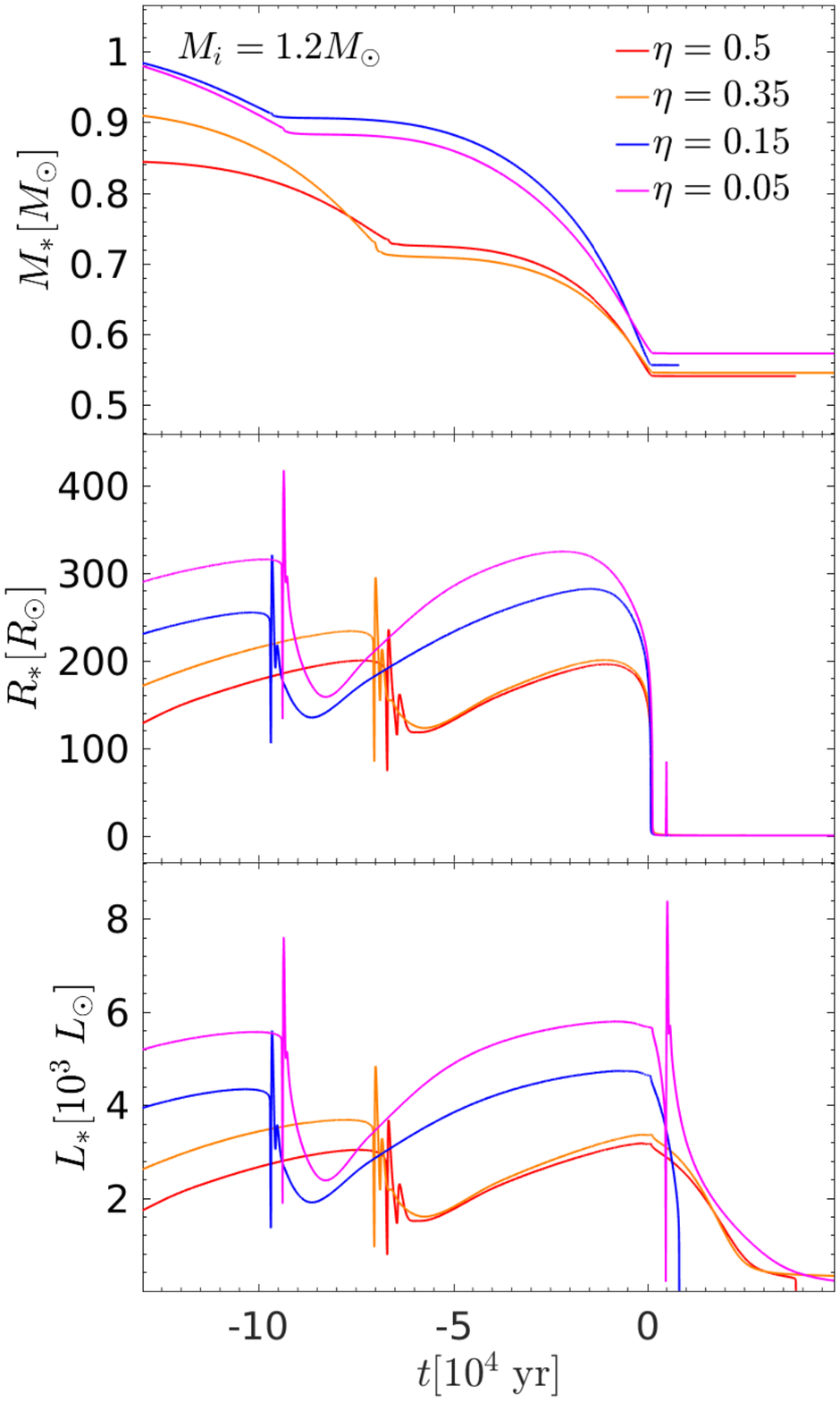}
\caption
{As in Figure \ref{fig:AGB_1M} but for an $M_i=1.2 M_\odot$ star.}
\label{fig:AGB_1.2M}
\end{figure}
\begin{figure}
\centering
\includegraphics
[trim= 3cm 1cm 4.6cm 1cm,clip=true,width=0.45\textwidth]
{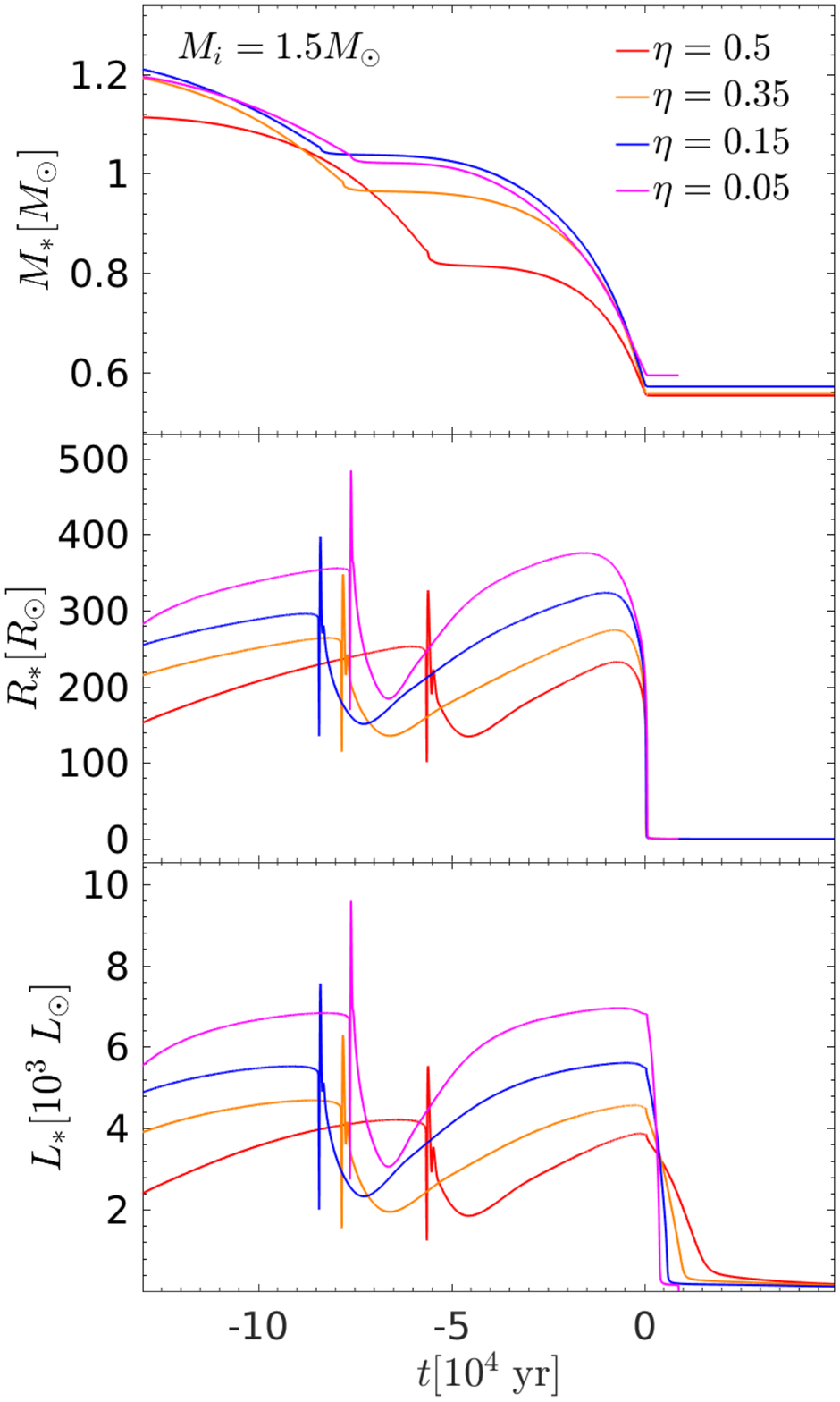}
\caption
{As in Figure \ref{fig:AGB_1M} but for an $M_i=1.5 M_\odot$ star.}
\label{fig:AGB_1.5M}
\end{figure}
\begin{figure}
\centering
\includegraphics
[trim= 3cm 1cm 4.6cm 1cm,clip=true,width=0.45\textwidth]
{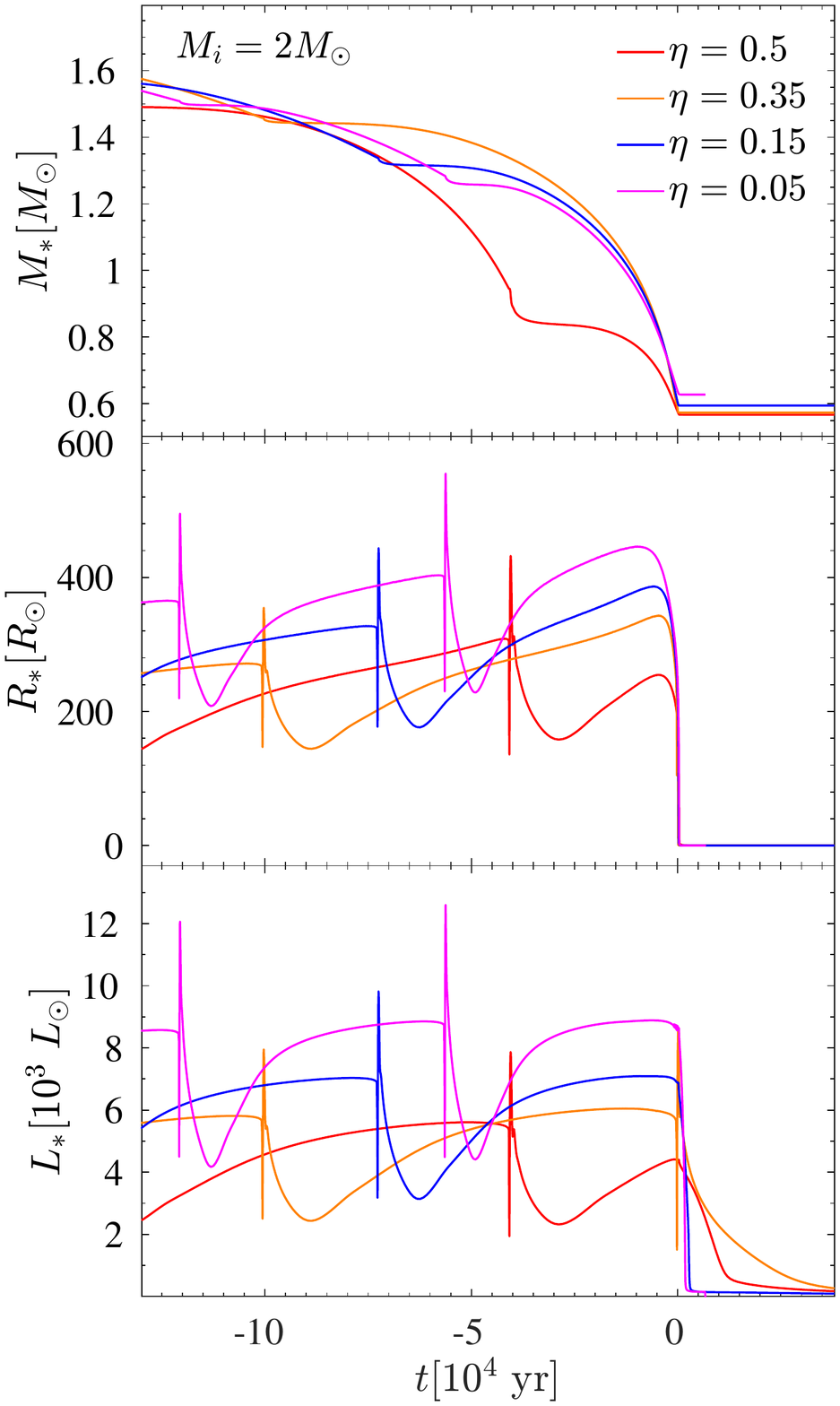}
\caption
{As in Figure \ref{fig:AGB_1M} but for an $M_i=2 M_\odot$ star.}
\label{fig:AGB_2M}
\end{figure}
\begin{table*}
    \centering          
    \begin{tabular}{lccccccc }
	\multicolumn{7}{ c }{Simulated models} \\	    
    \hline              
    \hline              
    $M_i$           &
    $\eta$          &
    $R_{\rm AGB}$   &
    $L_{\rm AGB}$   &
    $M_f$   		&
    $\xi_R$  		&
    $\xi_L$ 		&

\\  $[M_\odot]$ & &  $[R_\odot]$  &  $[L_\odot]$  &  $[M_\odot]$  &  & 
     \\[1ex]
    \hline              
    1   & 0.5   & 122 & $1.8\;\times10^{3}$ & 0.528  &  -   & -   
\\  1   & 0.35  & 174 & $2.5\;\times10^{3}$ & 0.534  & 1.4  & 1.4 
\\  1   & 0.25  & 199 & $3.0\;\times10^{3}$ & 0.540  & 1.6  & 1.7 
\\  1   & 0.15  & 229 & $3.6\;\times10^{3}$ & 0.546  & 1.9  & 2.0 
\\  1   & 0.07  & 272 & $4.4\;\times10^{3}$ & 0.554  & 2.2  & 2.4 
\\  1   & 0.05  & 288 & $4.9\;\times10^{3}$ & 0.560  & 2.4  & 2.7
\\ \hline 
\\  1.2 & 0.5   & 199 & $3.0\;\times10^{3}$ & 0.540  &  -   & -    
\\  1.2 & 0.35  & 230 & $3.6\;\times10^{3}$ & 0.545  & 1.2  & 1.2 
\\  1.2 & 0.25  & 234 & $3.7\;\times10^{3}$ & 0.549  & 1.2  & 1.2 
\\  1.2 & 0.15  & 280 & $4.7\;\times10^{3}$ & 0.559  & 1.4  & 1.6 
\\  1.2 & 0.07  & 313 & $5.4\;\times10^{3}$ & 0.567  & 1.6  & 1.8 
\\  1.2 & 0.05  & 324 & $5.8\;\times10^{3}$ & 0.572  & 1.6  & 1.9 
\\ \hline 
\\  1.5 & 0.5   & 250 & $4.2\;\times10^{3}$ & 0.552  &  -   & -    
\\  1.5 & 0.35  & 270 & $4.6\;\times10^{3}$ & 0.557  & 1.1  & 1.1 
\\  1.5 & 0.25  & 300 & $5.1\;\times10^{3}$ & 0.562  & 1.2  & 1.2 
\\  1.5 & 0.15  & 320 & $5.6\;\times10^{3}$ & 0.571  & 1.3  & 1.3 
\\  1.5 & 0.07  & 340 & $6.4\;\times10^{3}$ & 0.586  & 1.4  & 1.5 
\\  1.5 & 0.05  & 360 & $6.9\;\times10^{3}$ & 0.593  & 1.4  & 1.6 
\\ \hline 
\\  2   & 0.5   & 300 & $5.6\;\times10^{3}$ & 0.567  &   -  & -    
\\  2   & 0.35  & 320 & $6.0\;\times10^{3}$ & 0.573  & 1.1  & 1.1 
\\  2   & 0.25  & 340 & $6.2\;\times10^{3}$ & 0.582  & 1.2  & 1.1 
\\  2   & 0.15  & 380 & $7.0\;\times10^{3}$ & 0.595  & 1.3  & 1.2 
\\  2   & 0.07  & 400 & $8.3\;\times10^{3}$ & 0.616  & 1.3  & 1.4 
\\  2   & 0.05  & 420 & $8.8\;\times10^{3}$ & 0.627  & 1.4  & 1.6 
\\  [4ex]     
    \end{tabular}
\flushleft
\caption{ 
Results of our evolution simulations for solar metallicity ($Z=0.02$) stars:
the initial mass $M_i$, mass loss efficiency coefficient $\eta$ according to the Reimers mass loss equation (eq.\ref{eq:Reimers}), radius $R_{\rm AGB}$ and luminosity $L_{\rm AGB}$ at the last AGB pulses, and final mass of remnant $M_f$.
The radius and luminosity vary non-monotonically along the AGB,
so the value of $R_{\rm AGB}$ and $L_{\rm AGB}$ are approximate values taken during the last AGB pulses (not taking the He thermal pulses into account).
We also define the ratios of maximum AGB radii and luminosities,
between the Jsolated (reduced mass loss efficiency) and non-Jsolated stars (commonly used mass loss efficiency parameter of $0.5$) for each initial mass,
$\xi_R \equiv \frac{R_{\rm ABG} (\eta)}{R_{\rm ABG} (0.5)}$,
and $\xi_L \equiv \frac{L_{\rm ABG}  (\eta)}{L_{\rm ABG} (0.5)}$,
respectively, and list them in the last two columns.
}
\label{tab:results}
\end{table*}

From Figures \ref{fig:evolution} - \ref{fig:AGB_2M} and the properties we list in Table
\ref{tab:results} we conclude the following.
(1) The reduced mass loss rate we assume here for Jsolated stars, $\eta\la0.1$, brings
the radius on the AGB to be much larger
in comparison with non-Jsolated stars (commonly used mass loss rate).
Most significant is that the AGB radius of Jsolated stars is much larger
than their radius on the tip of the RGB.
(2) The AGB luminosities are much larger in cases with a reduced mass loss rate.  
(3) The final mass of the star, the bare AGB core, is larger for the reduced
mass loss rate cases, explaining the much larger post-AGB luminosities. 
 
\section{DISCUSSION AND SUMMARY}
\label{sec:discussion&summary}
In the present study we classified evolved stars that do not acquire much angular momentum, as expressed in eq. (\ref{eq:jsolated}), into a class that we term {\it Jsolated stars}. 
We suggested that the average mass loss rate of Jsolated stars is much
lower, by a factor of about five and more, than that of interacting starts, and followed the evolution of four stellar models from their zero age main sequence to the white dwarf phase.
We examined the evolution with several values of the mass loss rate efficiency parameter (eq. \ref{eq:Reimers}), ranging from the commonly used value of $\eta=0.5$ down to $\eta =0.05$. 
We presented the results in Figures \ref{fig:evolution}-\ref{fig:AGB_2M}, and in Table \ref{tab:results}.

We here focus on the effects of a reduced mass loss rate on the final AGB phase of low mass stars and the Planetary Nebula Luminosity Function (PNLF). We found that Jsolated stars with an initial mass of $1-2M_\odot$ reach much larger radii on their upper AGB and much higher luminosities on their upper AGB and post-AGB phases than non-Jsolated stars do.
The much larger radii implies that Jsolated stars can swallow a low mass
companion (either a low-mass main sequence star, or a brown dwarf, or a massive planet) and by that become non-Jsolated stars.
This interaction leads to the ejection of a dense nebula of a mass of
$\approx 0.2 M_\odot$.
The much brighter core then powers the bright [O~III]~$\lambda$5007 emission.

As evident from the fourth column of Table \ref{tab:results}, stars with initial masses of $M_{\rm i}= 1 M_\odot$ will only reach $L_{\rm pAGB} \approx 5000 L_\odot$ with a very low mass loss parameter of $\eta=0.05$, yet for the more massive cases Jsolated stars with
$\eta\la0.15$ reach high enough luminosities.
This, together with the interaction with a low mass companion on the upper AGB, will have the necessary ingredient for a bright PN, a dense nebula of mass $\ga 0.2 M_\odot$ and a bright central ionizing star. 

Some new stellar evolution calculations (e.g., \citealt{Karakas2014, Bertolami2016, Venturaetal2018}) find the post-AGB luminosity values  to be higher than in older previous calculations. 
We raise the possibility that these calculations together with our assumption of a very low mass loss rate of Jsolated stars with late engulfment of a very low mass companion, can account for the brightest PNe in old stellar populations. This combination, of higher AGB luminosities from new stellar simulations and higher luminosities from lower mass loss rates as proposed here, might actually be the most promising explanation to the puzzle.

\emph{The fate of the Earth.} 
In the commonly used mass loss rate, the fate of the Earth mostly
depends on the strength of the tidal interaction between Earth and the
giant Sun.
Due to the large sensitivity to tidal interaction (and even to
external planets; see \citealt{Veras2016}), different studies have
reached different conclusions on the question of whether the Sun will
swallow the Earth, maybe already during its RGB peak \citep{SchroderConnonSmith2008},
or whether the Earth will marginally survive engulfment
(e.g., \citealt{RybickiDenis2001}).
We note that the Sun is a Jsolated star,
or, in case it will swallow Jupiter, it will become a non-Jsolated star
only during the later stages of its AGB phase.
Whether the Sun will swallow Jupiter or not depends strongly on the poorly
known strength of the tidal interaction.
In Figure \ref{fig:evolution_1M} we present the ratio of the radius of a
solar model to the orbital separation of Earth.
Clearly, our assumption of a much lower mass loss rate, $\eta\la0.1$, of Jsolated stars
implies that Earth will be swallowed by the Sun and will be evaporated
in the giant envelope of the Sun, about 7 billion years from present day.
This conclusion does not depend on the tidal interaction strength.
\begin{figure}
\centering
\includegraphics
[trim= 3cm 6cm 4cm 4cm,clip=true,width=0.45\textwidth]
{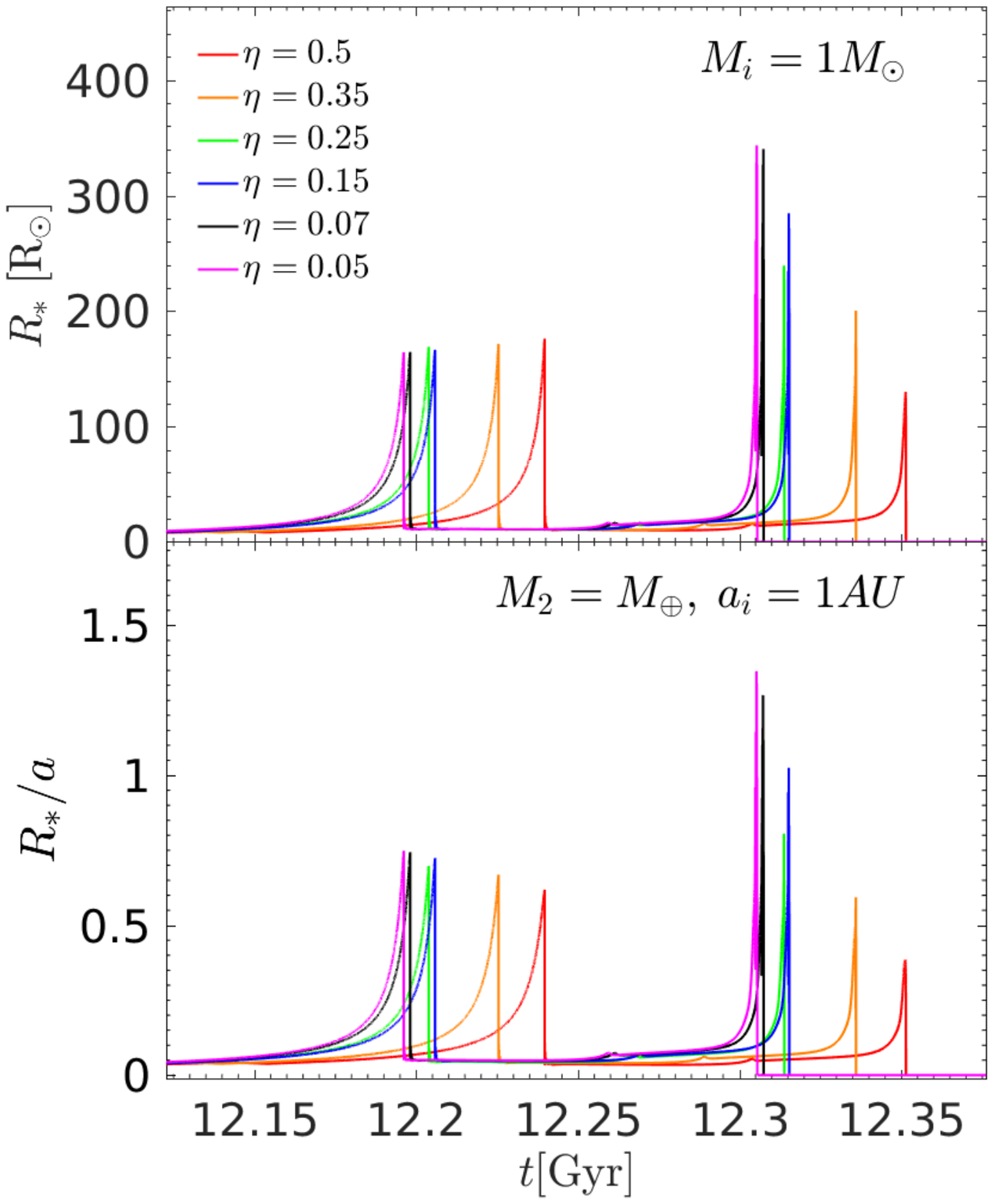}
\caption
{The radius of the Sun (upper panel), and the ratio of the evolving solar radius to the orbital separation
of Earth $R_\ast/a$ (lower panel), under the assumption of a reduced mass loss rate efficiency 
parameter.
The results are presented for 
$\eta=0.5$ (red), $\eta=0.35$ (orange),$\eta=0.25$ (green), $\eta=0.15$ (blue), $\eta=0.07$ (black), and $\eta=0.05$ (magenta).
}
\label{fig:evolution_1M}
\end{figure}

\section*{Acknowledgments}
We thank an anonymous referee for very helpful, detailed and thoughtful comments that improved the manuscript.
We acknowledge support from the Israel Science Foundation and a grant from the Asher Space Research Institute at the Technion.



\label{lastpage}
\end{document} 

---